# Magnetism of BaFe$_2$Se$_3$ studied by Mössbauer spectroscopy

K. Komędera[1], A. K. Jasek[1], A. Błachowski[1], K. Ruebenbauer[1*], M. Piskorz[2], J. Żukrowski[2,3], A. Krztoń-Maziopa[4], E. Pomjakushina[5], and K. Conder[5]

[1]Mössbauer Spectroscopy Division, Institute of Physics, Pedagogical University
*ul. Podchorążych 2, PL-30-084 Kraków, Poland*

[2]AGH University of Science and Technology,
Faculty of Physics and Applied Computer Science
*Av. A. Mickiewicza 30, PL-30-059 Kraków, Poland*

[3]AGH University of Science and Technology,
Academic Center for Materials and Nanotechnology
*Av. A. Mickiewicza 30, PL-30-059 Kraków, Poland*

[4]Warsaw University of Technology, Faculty of Chemistry
*ul. Noakowskiego 3, PL-00-664 Warsaw, Poland*

[5]Laboratory for Developments and Methods, Paul Scherrer Institut
*CH-5232 Villigen PSI, Switzerland*

[*]Corresponding author: sfrueben@cyf-kr.edu.pl

**Abstract**

The compound BaFe$_2$Se$_3$ (*Pnma*) has been synthesized in the form of single crystals with the average composition Ba$_{0.992}$Fe$_{1.998}$Se$_3$. The Mössbauer spectroscopy used for investigation of the valence states of Fe in this compound at temperature ranging from 4.2 K till room temperature revealed the occurrence of mixed-valence state for iron. The spectrum is characterized by sharply defined electric quadrupole doublet above magnetic ordering at about 250 K. For the magnetically ordered state one sees four iron sites at least and each of them is described by separate axially symmetric electric field gradient tensor with the principal component making some angle with the hyperfine magnetic field. They form two groups occurring in equal abundances. It is likely that each group belongs to separate spin ladder with various tilts of the FeSe$_4$ tetrahedral units along the ladder. Two impurity phases are found, i.e., superconducting FeSe and some other unidentified iron-bearing phase being magnetically disordered above 80 K. Powder form of BaFe$_2$Se$_3$ is unstable in contact with the air and decomposes slowly to this unidentified phase exhibiting almost the same quadrupole doublet as BaFe$_2$Se$_3$ above magnetic transition temperature.

## 1. Introduction

Iron-selenium compounds attracted recently significant attention as the iron-selenium bonds play important role in establishing superconductivity in numerous iron-based superconductors [1]. Generally, one observes that metallic iron-selenium compounds are prone to the loss of iron magnetic moment and therefore some of them can exhibit superconductivity at low temperature [2]. On the other hand, iron-selenium compounds being semiconductors exhibit iron magnetic moment characteristic of the $Fe^{2+}$ configuration – typically of the $S=2$ configuration. Mixed-valence configurations are encountered as well. Spin structures could be quite complex like e.g. spin ladders in the $BaFe_2Se_2O$ compound [3]. The compound $BaFe_2Se_3$ crystallizes in the orthorhombic structure with the *Pnma* symmetry. The structure is not very stable as application of the pressure of about 60 kbar at room temperature leads to the higher *Cmcm* symmetry. A reversible first-order iso-structural transition is observed at about 425 K and at ambient pressure as well. One observes further second order transition to the higher symmetry *Cmcm* at about 660 K and at ambient pressure [4]. A rotation of the $FeSe_4$ tetrahedral units is observed versus temperature at low temperatures up to about 340 K, while a displacement of Ba atoms dominates structural changes versus temperature above this temperature till transition to the *Cmcm* structure [4]. Hence, the crystallographic properties of this compound are highly unusual indicating a competition between various interactions. The magnetic order is unusual, too. Ferromagnetic $Fe_4$ blocks are formed at relatively high temperature and they order antiferromagnetically at about 255 K with magnetic moment per iron atom amounting to about $2.8\,\mu_B$ close to the ground state [5]. The long-range magnetic order established below 255 K is characterized by presence of two spin ladders within the system and two $Fe_4$ blocks per chemical cell [6]. The compound is found to be multi-ferroic at low temperature with the ferrielectric type of order. The net electric polarization for the lowest energy-state is expected to be almost negligible due to the cancellation between two ladders composed of the $FeSe_4$ tetrahedral units. However, this is not the case due to the presence of the strong magneto-elastic effects, i.e. magnetostriction [6]. The electronic structure is affected by the long-range magnetic order and it shows contributions from the localized and band states in the long-range magnetically ordered state [7]. No superconductivity was detected till practically the ground state of the system [8]. Single crystals were successfully synthesized, albeit all samples contain some spurious amounts of the FeSe superconductor [9]. Magnetic measurements on single crystals have been performed versus temperature as well confirming data obtained by other methods [10]. Band structure calculations for the *Pnma* form and within the LDA/LMTO approximation are reported, too [11]. These calculations show that antiferromagnetic coupling between ferromagnetic $Fe_4$ blocks leads to the lowest energy-state of the system. Early Mössbauer work was performed for $BaFe_2S_3$ (*Cmcm*) by using 14.41-keV resonant transition in $^{57}$Fe. Mixed-valence was found for iron. The spectrum was composed of a quadrupole-split doublet [12]. Further work [13] revealed single magnetically split site for the $BaFe_2S_3$ (*Cmcm*) compound and complex multiple site magnetically split spectra for the $BaFe_2Se_3$ (*Pnma*) compound. A magnetic ordering temperature was found to be at about 100 K for both compounds. This finding led to some speculations about spin fluctuation in $BaFe_2Se_3$ due to the discrepancy in the magnetic ordering temperatures found by the Mössbauer spectroscopy and neutron scattering [5]. Related compound $BaFe_2Se_2O$ has been investigated by Mössbauer spectroscopy as well [3]. The electric field gradient (EFG) tensor acting on the iron nucleus was found in the magnetically ordered state (close to the ground state) with extremely large anisotropy parameter, and with the hyperfine field making some undetermined (and maybe variable) angle with the principal component of the EFG. The situation has some similarity to the situation in otherwise metallic systems like FeAs [14] and FeSb [15], where interplay between covalent and itinerant electrons including significant orbital contributions is important. Recent work [16] suggests that $BaFe_2Se_3$ compound exhibits properties of the orbital-selective Mott phase.

Therefore, the compound $BaFe_2Se_3$ deserves reinvestigation by the Mössbauer spectroscopy versus temperature especially in the magnetically ordered region.

## 2. Experimental

$BaFe_2Se_3$ single crystals were prepared by the Bridgman method. To this end the high purity (at least 4N) powders of iron and selenium were combined in a molar ratio 2:3 and pressed into pellets of 5 mm-diameter. The prepared pellets were then sealed in a double-wall quartz ampoule together with a stoichiometric amount of metallic barium. The ampoule was heated to 1150 °C over 3 h and left at this temperature over 24 h to achieve suitable homogenization of the melt. Then the melt has been slowly (6 °C/h) cooled down to 750 °C followed by further annealing at this temperature for 20 h. Afterwards the material was cooled down to room temperature at the rate 200 °C/h. The method applied allowed for preparation of relatively brittle black crystallites, which cleave easily into plates.

Crystal structure of $BaFe_2Se_3$ sample was determined by means of the powder X-ray diffraction method. Measurements were performed at room temperature with D8 Advance Bruker AXS diffractometer using Cu $K_\alpha$ radiation (1.5406 Å). Measurements were done on crystal powdered in an inert atmosphere and loaded into the low background airtight sample holder to protect the material from oxidation. Reflections shown in the X-ray pattern (Figure 1) were indexed within the orthorhombic crystal system (space group *Pnma*). The lattice constants were found as $a$ = 11.9466 Å, $b$ = 5.4308 Å and $c$ = 9.1705 Å. The refinement of the crystal structure parameters was performed applying the FULLPROF application with the use of its internal tables [17].

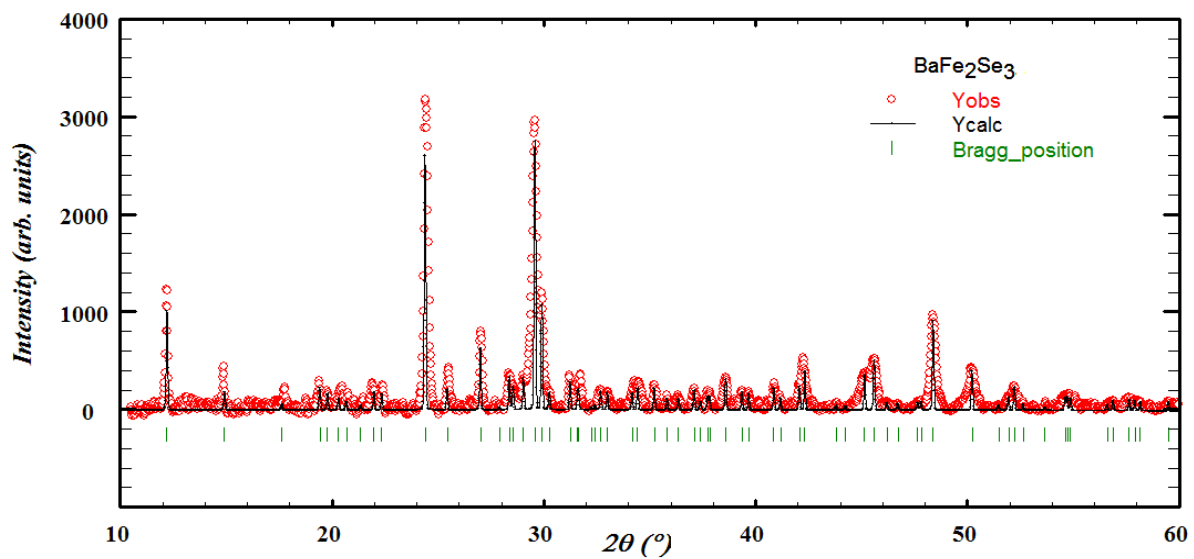

Fig. 1 Powder X-ray pattern of the $BaFe_2Se_3$ compound obtained at room temperature using Cu $K_\alpha$ radiation (1.5406 Å). The symbol $2\theta$ denotes scattering angle.

The X-ray fluorescence method applying micro-beam was used to determine composition maps of the sample. The average composition amounts to $Ba_{0.992}Fe_{1.998}Se_3$.

The absorber for Mössbauer spectroscopy was made in the powder form mixing 33 mg of the material with graphite and pressing into pellet having 14 mm diameter. In order to avoid oxidation the pellet was covered by the triple fluoropolymer layer. The absorber was made in above form and packed into silica ampoule (used for transportation) within the glove box filled with argon. Mössbauer spectra were obtained in the transmission geometry with the help of the Renon MsAa-4 spectrometer. A commercial $^{57}Co(Rh)$ source kept at ambient conditions was used. Sample temperature was set and controlled by using ICE helium cryostat. Spectra were fitted within transmission integral approximation by means of the Mosgraf-2009 suite [18]. Mössbauer spectra were recorded in two subsequent series. Spectra of the series 1 were collected in the order: at room temperature and 4.2 K. Spectra of the following series 2 were obtained in the temperature range 80 – 270 K with the increasing temperature. A time period between two series was two months and the epoxy resin layer was slightly damaged due to the thermal shock experienced during series 1. Additional spectrum was obtained at 80 K one month after series 1.

## 3. Results and discussion

Mössbauer spectra of series 1 are shown in Figure 2. Sub-spectra are shown as well (this statement applies to subsequent Figure 3, too). They are numbered in the same order as respective contributions to the total cross-section in Table 1. One has to bear in mind that the sum of sub-spectra areas is slightly greater than respective total spectral area due to the saturation effect. However this effect is small for spectra in question as even at 4.2 K the effective dimensionless absorber thickness was about 0.23, and slightly less at higher temperatures, i.e., about 0.19 at room temperature. The room temperature spectrum consists of two quadrupole doublets. A doublet showing minor contribution (6 % of the total cross-section) is due to the spurious phases (see, Table 1 for details). The major doublet exhibits rather narrow lines indicating unique iron site in the magnetically disordered $BaFe_2Se_3$ compound. A total spectral shift and quadrupole splitting confirm that iron stays in the mixed-valence state [12]. The spectrum obtained at 4.2 K could be fitted by six different iron sites. Two of them belong to the minor phases with the total contribution of about 7 %. One of these phases (3 %) is superconducting FeSe [2]. Hence, there are four different iron sites in magnetically ordered $BaFe_2Se_3$. Each of them could be fitted by the axially symmetric electric field gradient tensor (EFG) with the principal axis making some polar angle with the hyperfine magnetic field. They form two groups with approximately the same abundance. The group with the larger magnetic field has slightly larger shift (lower electron density on the nucleus) in comparison with the second group. The quadrupole coupling constant is larger within this group while the polar angle is larger as well and closer to the magic angle of the first order approximation – here the full Hamiltonians were used to process data. Presence of above two groups is likely to be caused by presence of two spin ladders within the $BaFe_2Se_3$ structure. A difference in the polar angle between groups is about 30º. Hence, this difference might be caused by rotation of the $FeSe_4$ tetrahedral units with lowering of the temperature in different way for each ladder. One can see two different iron sites in each ladder at least. The sites are similar one to another and generally one can expect more than two sites per ladder. Hence, relative abundances of two sites within a ladder cannot be taken very seriously as e.g. one sees average over four sites building the $Fe_4$ block and the hyperfine interactions are described by the non-diagonal Hamiltonians. A small difference in the polar angle (about 10º) might be an indication of the twist between $FeSe_4$ tetrahedral units along the ladder.

Spectrum obtained at 80 K one month since the series 1 and spectra of the series 2 are shown in Figure 3. They are characterized by increasing amount of the unidentified phase amounting to 4 % in the 4.2 K spectrum of series 1. A contribution due to this phase masks eventually contribution of the superconducting FeSe. This phase has very small hyperfine magnetic field of about 5.2 T at 4.2 K and it becomes magnetically disordered above 80 K. It is interesting to note that the hyperfine pattern of this phase is hard to distinguish from the pattern of $BaFe_2Se_3$ at room temperature. Despite decomposition of the sample one observes a contribution to the spectrum due to unperturbed $BaFe_2Se_3$ with a magnetic transition at about 250 K in accordance with the neutron scattering data [5].

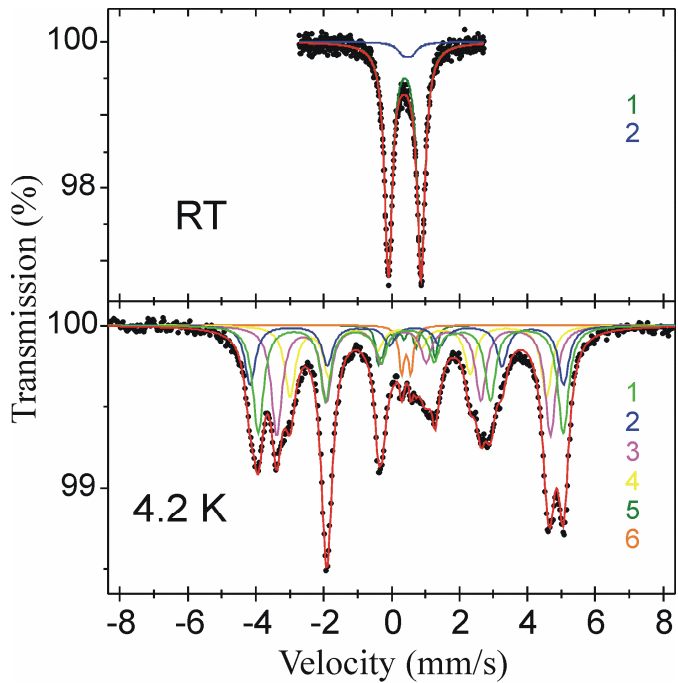

Fig. 2 $^{57}$Fe Mössbauer spectra belonging to the series 1. Sub-spectra are numbered in the same order as respective contributions in Table 1.

Fig. 3 $^{57}$Fe Mössbauer spectrum obtained between series (top) and spectra belonging to series 2. Sub-spectra are numbered in the same order as respective contributions in Table 1.

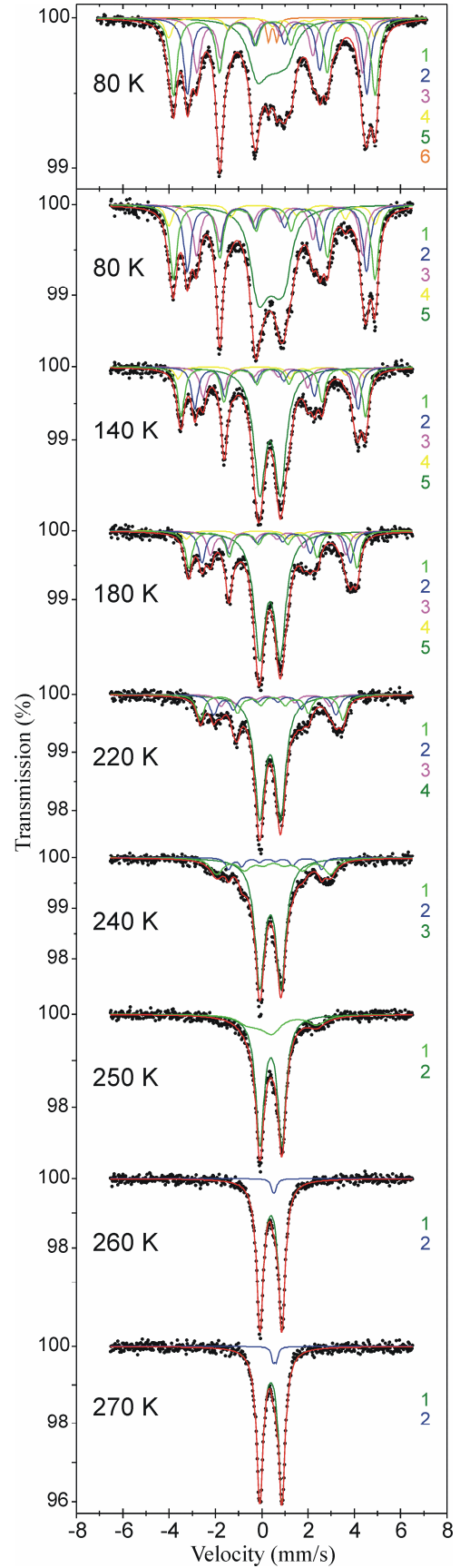

**Table 1**

Essential results obtained by Mössbauer spectroscopy. The symbol $T$ stands for the absorber temperature (RT – denotes room temperature). The symbol $C$ denotes relative contribution of the particular component, while the symbol $S$ stands for a total shift versus room temperature α-Fe. The symbol $A_Q$ stands for the quadrupole coupling constant including sign for magnetically split components, while the symbol $\Delta$ denotes quadrupole splitting of the components without hyperfine magnetic field. The symbol $B$ stands for the hyperfine magnetic field, while the symbol $\theta$ denotes angle between principal component of the electric field gradient and hyperfine magnetic field. The symbol $\Gamma$ stands for the absorber line width.

| $T$ (K) | $C$ (%) | $S$ (mm/s) | $6A_Q$ (mm/s) $\Delta$ (mm/s) | $B$ (T) | $\theta$ (°) | $\Gamma$ (mm/s) |
|---|---|---|---|---|---|---|
| | | | series 1 | | | |
| RT | 94 | 0.504 | 0.958 | - | - | 0.18 |
| | 6 | 0.59 | 0.24 | - | - | 0.3 |
| 4.2 | 29 | 0.64 | 1.14 | 27.0 | 51 | 0.28 |
| | 16 | 0.68 | 1.14 | 27.9 | 64 | |
| | 29 | 0.62 | 0.60 | 24.7 | 35 | |
| | 19 | 0.62 | 0.84 | 23.0 | 27 | |
| | 4 | 0.46 | 0.24 | 5.2 | - | 0.1 |
| | 3 | 0.55 | 0.27 | - | - | 0.1 |
| | | | one month after series 1 | | | |
| 80 | 26 | 0.64 | 1.26 | 25.9 | 53 | 0.28 |
| | 25 | 0.62 | 0.66 | 23.7 | 35 | |
| | 18 | 0.62 | 0.84 | 21.9 | 24 | |
| | 7 | 0.79 | 2.76 | 23.6 | 68 | |
| | 22 | 0.47 | 0.18 | 4.3 | - | 0.7 |
| | 2 | 0.58 | 0.36 | - | - | 0.1 |
| | | | series 2 – two months after series 1 | | | |
| 80 | 24 | 0.65 | 1.26 | 25.9 | 54 | 0.27 |
| | 21 | 0.62 | 0.60 | 23.7 | 34 | |
| | 16 | 0.61 | 0.78 | 22.0 | 23 | |
| | 7 | 0.90 | 2.70 | 24.2 | 73 | |
| | 32 | 0.49 | 0.06 | 3.9 | - | 0.62 |
| 180 | 19 | 0.61 | 1.20 | 21.3 | 55 | 0.26 |
| | 16 | 0.61 | 0.72 | 19.4 | 39 | |
| | 12 | 0.56 | 0.66 | 18.0 | 19 | |
| | 4 | 0.90 | 2.76 | 18.8 | 79 | |
| | 49 | 0.47 | 0.89 | - | - | 0.48 |
| 240 | 23 | 0.63 | 0.90 | 14.3 | 55 | 0.53 |
| | 8 | 0.53 | 0.48 | 12.2 | 25 | 0.20 |
| | 69 | 0.50 | 0.91 | - | - | 0.40 |
| 270 | 96 | 0.509 | 0.951 | - | - | 0.25 |
| | 4 | 0.68 | 0.14 | - | - | 0.1 |

## 4. Conclusions

In general, Mössbauer spectroscopy sees the same long-range magnetic ordering temperature as other methods including neutron scattering. A discrepancy in magnetic ordering temperatures seen before [13] is due to the instability of the powder form of this compound in the contact with the air at ambient temperature. A decomposition product might be related to the recently discovered phase of $BaFe_2Se_3$ formed at high-pressure [4]. The Mössbauer spectra of the decomposition product and the $BaFe_2Se_3$ phase are almost the same above the long-range magnetic ordering temperature of $BaFe_2Se_3$. It seems that magnetic ordering of $BaFe_2Se_3$ leads to some atomic position variation, as the EFG tensors observed at low temperature seem to look like due to the differentiation of the unique EFG seen above magnetic transition. Presence of two groups of iron sites seen in the magnetically ordered state is likely to reflect presence of two spin ladders within the structure. Different angles between principal axis of EFG and hyperfine field for various iron sites confirm various tilts of the $FeSe_4$ tetrahedral units along spin ladders.

**Acknowledgments**

This work was supported by the National Science Center of Poland, Grant No. DEC-2011/03/B/ST3/00446. A.K.-M. acknowledges the financial support founded by the National Science Center of Poland, Grant No. DEC-2013/09/B/ST5/03391.